\documentclass{nature}

\usepackage{graphicx}
\usepackage{amsmath,amssymb}

\title{Nanometer scale quantum thermometry in a living cell}

\author{G. Kucsko$^{1 *}$, P. C. Maurer$^{1 *}$, N. Y. Yao$^{1}$, M. Kubo$^{2}$, H. J. Noh$^{3}$,  P. K. Lo$^{4}$, H. Park$^{1,2,3}$, and  M. D. Lukin$^{1}$}

\begin{document}

\maketitle

\begin{affiliations}
 \item Department of Physics, Harvard University, Cambridge, MA 02138, USA.
 \item Department of Chemistry and Chemical Biology, Harvard University, Cambridge, MA 02138, USA.
 \item Broad Institute of MIT and Harvard, 7 Cambridge Center, Cambridge, MA, 02142, USA.
 \item Department of Biology and Chemistry, City University of Hong Kong, Hong Kong, Kowloon, China.
 \item[$^{*}$] These authors contributed equally to this work

\end{affiliations}

\begin{abstract}
Sensitive probing of temperature variations on nanometer scales represents an outstanding challenge in many areas of modern science and technology\cite{yue2012nanoscale}.
In particular, a thermometer capable of sub-degree temperature resolution as well as integration within a living system could provide a powerful new tool for many areas of biological research, including temperature-induced control of gene expression\cite{lucchetta2005dynamics, kumar2010h2a, lauschke2012scaling, kamei2008infrared} and cell-selective treatment of disease\cite{schroeder2011treating,lee2011exchange,o2004photo}.
Here, we demonstrate a new approach to nanoscale thermometry that utilizes coherent manipulation of the electronic spin associated with nitrogen-vacancy (NV) color centers in diamond.
We show the ability to detect temperature variations down to \textmd{1.8~mK} (sensitivity of \textmd{$9~\text{mK} / \sqrt{\text{Hz}}$}) in an ultra-pure bulk diamond sample.
Using  NV centers in diamond nanocrystals (nanodiamonds), we directly measure the local thermal environment at length scales down to $200$~\textmd{nm}.
Finally, by introducing both nanodiamonds and gold nanoparticles into a single human embryonic fibroblast, we demonstrate temperature-gradient control and mapping at the sub-cellular level,  enabling unique potential applications in life sciences.
\end{abstract}

The ability to monitor sub-kelvin variations over a large range of temperatures can provide insight into both organic and inorganic systems, shedding light on questions ranging from tumor metabolism\cite{vreugdenburg2013systematic} to heat dissipation in integrated circuits\cite{yue2012nanoscale}.
Moreover, by combining local light-induced heat sources with sensitive nanoscale thermometry, it may be possible to engineer biological processes at the sub-cellular level.
 Many promising approaches are currently being explored for this purpose, including scanning probe microscopy\cite{yue2012nanoscale,majumdar1999scanning}, Raman spectroscopy\cite{kim2006micro}, and fluorescence-based measurements using nanoparticles\cite{yang2011quantum, vetrone2010temperature} and organic dyes\cite{okabe2012intracellular, donner2012mapping}.
These methods, however, are often limited by a combination of low sensitivity, bio-incompatibility, or systematic errors owing to changes in the local chemical environment.

\indent Our new approach to nanoscale thermometry utilizes the quantum mechanical spin associated with  nitrogen vacancy (NV) color centers in diamond. The operational principle of NV-based thermometry relies upon the temperature dependent lattice strain of diamond\cite{acosta2010temperature}; changes in the lattice are directly reflected as changes in the spin properties of the NV, which are then optically detected with high spatial resolution.
As illustrated in Fig.~1, each NV center constitutes a spin-1 degree of freedom with a zero-field splitting $\Delta$ between the $|m_s = 0 \rangle$ and the $|m_s = \pm1 \rangle$ levels.  This zero-field splitting exhibits a strong temperature dependence ($d\Delta / dT = -(2\pi) 77~\text{kHz} / \text{K}$ at $300~\text{K}$),  enabling fluorescence-based thermometry via precision electron spin resonance (ESR) spectroscopy\cite{acosta2010temperature,toyli2012measurement,chen2011temperature}.
For a sensor containing $N$ color centers, the temperature sensitivity is given by
\begin{equation}
\eta =  \frac{C} {d\Delta / dT} \frac{1}{\sqrt{T_{\text{coh}} Nt}},
\end{equation}
where $T_{\text{coh}}$ is the NV spin coherence time and $t$ is the integration time. Here, we also introduce a factor $C$ to account for imperfect readout and initialization\cite{taylor2008high}.
Assuming $T_{\text{coh}}$ on the order of few milliseconds and $C\approx0.03$\cite{taylor2008high}, a single NV can potentially exhibit a sensitivity better than  $1~\text{mK} / \sqrt{\text{Hz}}$.
Beyond high sensitivity, NV-based thermometry offers several distinct advantages over existing methods in biological and chemical temperature sensing.
First, owing to diamond's chemical inertness, it is generally robust to changes in the  local chemical environment.
Second, our method can be applied over a wide range of temperatures, $200-600$~K\cite{toyli2012measurement,chen2011temperature}.

As a first benchmarking experiment, we demonstrate the high temperature sensitivity of NV-based thermometry in a bulk diamond sample.
While the NV's magnetic sensitivity has rendered it a competitive magnetometer\cite{maze2008nanoscale, balasubramanian2008nanoscale}, to accurately determine the temperature, it is necessary to decouple the NV electronic spin from fluctuating external magnetic fields.
This is achieved via a modified spin-echo sequence that makes use of the spin-1 nature of the NV defect\cite{hodges2011time}, allowing us to eliminate the effects of an external, slowly varying, magnetic field.
Specifically, we apply microwave pulse at frequency $\omega$ (Fig.~1B) to create a coherent superposition, $\frac{1}{\sqrt{2}} ( |0\rangle + |B \rangle)$, where $|B\rangle = \frac{1}{\sqrt{2}} (|+1\rangle + |-1\rangle)$. After half the total evolution time, $\tau$, we apply a $2 \pi$ echo-pulse that swaps the $|+1 \rangle$ and $|-1\rangle$ states (Fig.~2A). Following another period of $\tau$ evolution, quasi-static (e.g. magnetic-field-induced) shifts of these $|\pm1 \rangle$ levels are eliminated, allowing for accurate temperature sensing. In the experiment, we use a CVD-grown, isotopically pure diamond ($99.99~\%$ spinless $^{12}$C isotope) sample\cite{balasubramanian2009ultralong} to further reduce magnetic-field fluctuations originating from the intrinsic $^{13}$C nuclear spin bath.
As shown in Fig.~2B, this  allows us to observe coherence fringes approaching $0.5~\text{ms}$.
Interestingly, for all NVs tested, we observe a characteristic  low-frequency beating of the fluorescence signal that varies from NV to NV, which is most likely due to  locally fluctuating charge traps\cite{dolde2011electric}. Despite this beating, we observe a temperature sensitivity $\eta = (9 \pm1.8)~\text{mK} / \sqrt{\text{Hz}}$ by recording the signal for a fixed evolution time $\tau = 250 ~\mu \text{s}$ (Fig.~2C). Within 30-second integration, we achieve a measurement accuracy $\delta T = 1.8 \pm 0.3$~mK.

We now demonstrate the high spatial resolution of NV-based thermometry, which can be achieved by using diamond nanocrystals (nanodiamonds).
In most commercially available nanodiamonds, the NV coherence time is  limited to approximately $1~\mu$s due to additional paramagnetic impurities. While this shortened coherence time reduces the intrinsic temperature sensitivity for a single defect, this decrease can be offset by using an ensemble of NVs to enhance the signal to noise ratio by a factor of $\sqrt{N}$.
Note that unlike NV-based magnetometry, where the proximity to the source (often limited by nanodiamond size) is critical to the maximum sensitivity, NV thermometry is not subject to such a constraint; in fact, the excellent thermal conductivity of diamond ensures that \emph{all} NV centers within a nanocrystal are in thermal equilibrium with the local heat environment.
To maximize the number of NV centers and to minimize the lattice strain, our measurements are performed on  single-crystalline nanodiamonds containing approximately $500$ NV centers (Adamas Nanotechnologies). The zero-field splitting $\Delta$ of the NV ensemble, and thus the temperature, is determined by recording a continuous-wave ESR spectrum. Specifically, we measure  changes to the zero-field splitting by recording the fluorescence at four different frequencies centered around $\Delta = 2.87$~GHz (Fig.~3A). This procedure eliminates unwanted effects from fluctuations in the total fluorescence rate, ESR contrast, Rabi frequency and magnetic field, yielding a robust thermometer (see Methods for details).

Combining our nanodiamond thermometer with the laser heating of a gold nanoparticle (Au NP) allows us to both control and monitor temperature at nanometer length scales (Fig.~3B,C).
Both nanodiamonds and Au NPs (nominal diameter $100$~nm) are initially spin-coated on a microscope coverslip. Using a home-built confocal microscope with two independent scanning beams, we co-localize an Au NP and a nanodiamond that are separated by $0.8\pm 0.1~\mu\text{m}$. While locally heating the Au NP via continuous illumination with a variable-power green laser (focused to a diffraction limited spot), we also simultaneously measure the temperature at the nanodiamond location using ESR spectroscopy.

Figure 3B shows the temperature change recorded by the nanodiamond as a function of the green laser power (red points). From a linear fit to the data we estimate the accuracy of our ND sensor to be $\delta T = (44\pm10)$~mK (see Methods for details).
The measured temperature change is in excellent agreement with the theoretically expected temperature profile based upon a steady-state solution of the heat equation, $\Delta T(r) = \frac{\dot{Q}}{4\pi \kappa r}$, where $\dot{Q}$ is the   heat dissipation, $\kappa$ is the thermal conductivity of glass and $r$ is the distance between the nanodiamond and the Au NP.
To further verify that the temperature change originates from the local heating of the Au NP, we repeat the measurement with the excitation laser displaced from the NP by $0.8$~$\mu$m. In this case, the temperature measured by the nanodiamond remained constant as a function of laser power (blue points), thereby confirming the locality of the heat source.
As shown in Fig.~3C,  we record the temperature of six nanodiamonds at different distances from the laser-heated Au NP;  we find that the measured temperature profile (Fig.~3D) as a function of distance is in excellent agreement with the theoretical steady-state prediction (solid line). This allows us to directly  estimate the temperature change at the location of the Au NP to be $72\pm6$~K.

To demonstrate that nanodiamond thermometry is compatible inside living cells, we introduce nanodiamonds and Au NPs into human embryonic fibroblast WS1 cells via nanowire-assisted delivery\cite{shalek2010vertical,ShalekDeliv}. Just as in the control experiments described above, we probe the temperature at two different locations (NV$_1$ and NV$_2$) within a single cell while locally heating an individual Au NP (Fig.~4A).
As shown in Fig.~4B, NV$_1$, which is significantly closer to the heat source, exhibits a stronger temperature dependence as a function of laser power than NV$_2$. Varying the incident  power allows us to generate controlled temperature gradients of up to $5$~K over distances of approximately $7~\mu$m.
To ensure that this temperature gradient is created by the controlled illumination of the NP and does not result from  heating of the cellular cytoplasm, we displace the laser spot from the Au NP; this then results in a negligible temperature change at the location of NV$_1$ with $\Delta T = (-20 \pm 50)$~mK (green square, Fig.~4B).
The increased measurement uncertainty for larger laser powers is the result of heating fluctuations from drift of the Au NP.

The experiments shown in Fig.~4B clearly demonstrate the sub-micron measurement of an intra-cellular heat gradient. However, the substantial heating induced by constant  illumination for an extended period of time, ultimately leads to the death of the cell, which is confirmed using a standard live/dead assay (Calcein AM/Ethidium Homodimer-1).
To demonstrate that our technique can be employed within living cells, we increase the concentration of Au NPs  to allow for heat generation at different locations by simply moving the laser focus.
Then, we measured the temperature variation at a single nanodiamond (bar plot in Fig.~4C) while introducing a slight heating of Au NPs in two differing locations (red, yellow crosses).
After our measurement, the viability of the cell is confirmed (Fig.~4C).

Finally, we demonstrate that our method can be used to control cell viability.
To start, we heat the cell with $12~\mu$W of laser power and measure a temperature change of $0.5\pm0.2$~K at the nanodiamond location; this corresponds to a change of approximately $10$K at the Au NP spot.
At this point, the cell is still alive, as confirmed by the absence of ethidium homodimer-1 fluorescence inside the  membrane (Fig.~4D).
By increasing the laser power to $120\mu$W, we induce a temperature change of $3.9\pm0.1$K at the nanodiamond location (approximately $80$K at the location of the laser focus); in this case, the cell is flooded with  fluorescence from the ethidium homodimer, thus signaling cell death.
 This proof-of-principle experiment indicates that nanodiamond thermometry may help enable the optimization of NP-based photothermal therapies\cite{pitsillides2003selective,o2004photo}.

Our experiments demonstrate that the quantum spin of NV centers in diamond can be used as a robust temperature sensor that combines the virtues of sub-micron spatial resolution, sub-degree thermal sensitivity and bio-compatibility.
The sensitivity of our current measurement can be enhanced by improving the relevant coherence time and by increasing the number of NV centers within the nanocrystal. Optimizing these factors should allow us to reach sensitivities of $80$~$\mu\text{K} / \sqrt{\text{Hz}}$ (Methods), yielding the ability to sense sub-kelvin temperature variations with milli-second time resolution. In solution, the  ultimate accuracy of our method will likely be limited by residual heating during the measurement process. As discussed in the Methods, this limit is in the range of $50~\mu$K to $5$~mK, depending on experimental conditions.  Furthermore, the spatial resolution of our method can  be  improved by using far-field sub-diffraction techniques\cite{maurer2010far}.

NV-based thermometry opens up a number of intriguing potential applications.
For instance, the simultaneous real-time measurement and control of a sub-cellular thermal gradient could enable the accurate control of gene expression\cite{xu2012identification}.
%As a near term experiment one might envision characterizing  sub-cellular heat response by observing the expression and localization of GFP-tagged temperature sensitive proteins\cite{xu2012identification} in response to optically regulated incandescence.
The large dynamic range of our quantum thermometer and it's intrinsic robustness may also allow for the direct microscopic monitoring and control of chemical reactions\cite{jin2011localized}.
Moreover, combining our technique with two-photon microscopy\cite{helmchen2005deep,wee2007two} may enable \emph{in vivo} identification of local tumor activity by mapping atypical thermogenesis at the single-cell level\cite{tsoli2012activation}.
Finally the combination of thermoablative therapy with our temperature sensor constitutes a potent tool for the selective identification and killing of malignant cells without damaging surrounding tissue\cite{schroeder2011treating,lee2011exchange,o2004photo}.

\begin{figure}

\caption{ {\bf Quantum spin-based thermometry }| \textbf{(A)} Schematic image depicting nanodiamonds and gold nanoparticles (Au NPs) within a living cell. The controlled application of local heat is achieved via laser illumination of the Au NP, while nanoscale thermometry is achieved via precision spectroscopy of the NV spins in nanodiamonds.  \textbf{(B)} Simplified NV level diagram showing a ground state spin triplet and an excited state. At zero magnetic field, the  $|\pm1\rangle$ sub-levels are split from the $|0 \rangle$ state by a temperature-dependent zero field splitting $\Delta (T)$. Pulsed microwave radiation is applied (detuning $\delta$) to perform Ramsey-type spectroscopy. \textbf{(C)} Comparison between the NV quantum thermometer and other reported techniques as a function of sensor size and temperature accuracy. Red circles indicate methods that are biologically compatible. The red open circle indicates the ultimate expected accuracy for our measurement technique in solution (see Methods).
}

\caption{ {\bf Sensitivity of Single NV Thermometer} | \textbf{(A)} Measured fluorescence as a function of echo evolution time $2\tau$ (red points); the black solid line indicates a fit corresponding to a damped cosine function with two distinct frequencies. The characteristic beating can be explained by  fluctuating proximal charge traps located at distances of about $50$~nm. The inset depicts the microwave $2\pi$-echo-pulse sequence used to cancel unwanted  external magnetic field fluctuations\cite{hodges2011time}.  \textbf{(B)} Measured fluorescence (red points), corresponding error bars (one standard deviation) and best fit line as function of temperature for an echo time $2 \tau= 250$~$\mu \textmd{s}$.}

\caption{{\bf Sub-micron thermometry using nanodiamonds} | \textbf{(A)} Frequency scan of a single nanodiamond containing approximately $500$ NV centers. The four red points indicate the measurement frequencies used to extract the temperature as detailed in Methods.  \textbf{(B)} Temperature of a single nanodiamond as a function of laser power for two different laser-focus locations. The red data points depict the dramatic  heating of a nanodiamond as a result of laser illumination on a nearby Au NP.  The blue data points depict the same measurement with the laser focus displaced by $0.8~\mu$m from the Au NP location; this results in the negligible heating of the nanodiamond as a function of laser power.  \textbf{(C)} Two-dimensional confocal scan of nanodiamonds (green circles) and Au NPs (yellow cross) spin-coated onto a glass coverslip. \textbf{(D)} Temperature changes measured (red points) at the six nanodiamond locations in (C) as a function of distance from the illuminated Au NP (yellow cross). The blue curve represents the theoretical temperature profile based upon a steady-state solution of the heat equation. All data in this figure are obtained on a glass coverslip, and all error bars correspond to one standard deviation.}

\caption{ {\bf Nanoscale thermometry in cells} | \textbf{(A)} Confocal scan of a single cell under $532$~nm excitation with collection above $638$~nm.
The white cross corresponds to the position of the Au NP used for heating, while red (NV$_1$) and blue (NV$_2$) circles represent the location  of nanodiamonds used for thermometry.
The dotted white line provides a guide to the eye and outlines the cell membrane.
\textbf{(B)} Measured change in temperature at the position of NV$_1$ and NV$_2$ relative to the incident  laser power applied to the Au NP. Dashed lines are linear fits to the data, and error bars represent one standard deviation.
\textbf{(C)} Fluorescence scan of stained cells (live/dead assay) with excitation at $494/528$~nm and emission at $515$~nm (green - cell alive) and $617$~nm (red - cell dead). The bar plot depicts the temperature of a single nanodiamond with local heat applied at two different locations (red/yellow cross).
\textbf{(D)} Confocal fluorescence scans of an individual cell under varying illumination power. Excitation occurs at $532$~nm and collection is above $630$~nm. Cell death is indicated by the penetration of ethidium homodimer-1 through the cell membrane, staining the nucleus. At low laser powers, the cell remains alive, while cell-death occurs as laser-induced heating is increased.  }

\end{figure}

% \section*{Another Section}

% \begin{methods summary}
\subsection{Methods Summary}

\subsection{Nanodiamond measurement pulse sequence |}
As indicated in figure 3A, we record the fluorescence at four different frequencies centered around $\Delta =2.87$GHz: \\*
$f^{1,2} \approx f\left( \omega_- \right) + \frac{\partial f}{\partial \omega} \vert_{\omega-} \left( \mp \delta\omega + \delta B + \delta T \frac{d\Delta}{dT} \right)$ and
$f^{3,4} \approx f\left( \omega_+ \right) + \frac{\partial f}{\partial \omega} \vert_{\omega+} \left( \mp \delta\omega - \delta B + \delta T \frac{d\Delta}{dT} \right)$. This allows us to determine the change in temperature,
\begin{equation}
\delta T = \frac{\delta \omega}{d\Delta / dT} \frac{\left( f^1+f^2 \right) - \left( f^3+f^4 \right)}{\left( f^1-f^2 \right) + \left( f^3-f^4 \right)},
\end{equation}
where $\omega_{\pm} \mp \delta \omega$ are the four microwave carrier frequencies and $\delta B$ is a  unknown static magnetic field. By averaging the fluorescence at these four frequencies  as shown in equation (2), we are able to remove errors associated with changes in total fluorescence rate, ESR contrast, Rabi frequency and magnetic field.
% For current experimental parameters we can estimate a sensitivity of $100~\text{mK} / \sqrt{\text{Hz}}$.

% \end{methods summary}

% \begin{methods}
\subsection{Methods}

\subsection{Experimental apparatus, sensitivity and accuracy |}
Our experimental apparatus consists of a confocal microscope with two independent excitation/collection paths allowing measurement and heating at two independent locations simultaneously.
The experiments use either a Nikon Plan Fluor 100x oil immersion, NA = 1.3, (nanodiamonds) or a Nikon Plan Apo 100x air, NA = 0.95, objective (bulk sample), resulting in $C \approx 0.03$, which can be further improved by employing a solid immersion lens or diamond nano patterning.
Microwaves are delivered via a lithographically defined coplanar waveguide on top of a glass coverslip.
For experiments with nanodiamonds we use neutral density filters in the collection path to avoid saturation of the APD.
The temperature accuracy $\delta T$ for bulk diamond is estimated from the measurement shown in Fig.~2B.
Using the standard deviation $\sigma$ (shown error bars) we evaluate the accuracy as $\delta T = \sigma / (c~\frac{d\Delta}{dT}~2 \tau)$, where $c$ is the oscillation amplitude and $2 \tau$ is the free evolution time.
We find that for integration times $t < 30~s$ (limited by temperature stability) the temperature accuracy improves as $\sqrt{t}$, giving a sensitivity $\eta = \delta T \sqrt{t}$.
A linear dependence of the dissipated heat as a function of laser power (Fig.~3B) is used to determine the measurement accuracy for nanodiamonds.
A linear function, with slope $m$, is fitted to the data (red dashed line) and the measurement accuracy is given by $\delta T = \sqrt{ \frac{1}{N-1} \Sigma_{i=1}^{N} \left( T_i - m~P_i \right)^2}$, with $T_i$ the measured temperature and $P_i$ the corresponding laser power.
The error bar is evaluated as $\sigma \left( \delta T \right) = \delta T \sqrt{1 - \frac{2}{N-1} \frac{ \Gamma^2 \left( n/2 \right) }{ \Gamma^2 \left( (n-1)/2 \right) }}$, where $\Gamma (\cdot)$ indicates the Gamma distribution.

\subsection{Ultimate sensitivity |}
The ultimate sensitivity of our method is limited by the NV coherence time and the number of defect centers. In our current experiment, we have demonstrated a sensitivity of $9$~$\text{mK} / \sqrt{\text{Hz}}$ (with a free evolution time of $250~\mu$s). Two natural extensions  enable longer NV coherences: 1) decreasing the $^{13}$C concentration to suppress the nuclear spin bath and 2) further dynamical decoupling. These methods can, in principle, allow us to extend the evolution time up to $T_1 / 2 \sim 3$~ms. In combination with a nanocrystal that contains $\sim 1000$ NV centers, this could yield a ultimate sensitivity limit as $80$~$\mu\text{K} / \sqrt{\text{Hz}}$. Further improvement may be possible by employing spin squeezed states~\cite{cappellaro2009quantum}.

\subsection{Ultimate accuracy in solution |}
In cases where our method is used to probe a system that is in solution (e.g. cells, chemical reactions), the primary accuracy limit is set by heat dissipation during the measurement process. In particular, the microwave spectroscopy used to detect changes in the NV zero field splitting also induces heating of the solution. In the present experiment, we utilize a lithographically fabricated microwave loop (diameter $200~\mu$m) to generate an ac-magnetic field, $B \approx 10$~milli-gauss, for spin manipulations. Estimating the effective dipole field created by the microwave loop shows that the solution (water) absorbs $10^{-6}$~W of power yielding a temperature increase of $5$~mK in the steady state.  By using a smaller microwave loop ($20~\mu$m) and reducing the duty cycle, it is possible to decrease the heating of the solution to approximately $50~\mu$K.

\subsection{Injection of nanodiamonds in cells  |}
Nanodiamonds and Au NPs were introduced into WS1 cells via silicon nanowire-mediated delivery\cite{shalek2010vertical}. Silicon nanowires were treated with 3-amino- propyltrimethoxysilane to present  ÐNH$_2$ functionality on the surface, and nanodiamonds / Au NPs were subsequently attached via electrostatic binding. Afterwards, human embryonic fibroblast WS1 cells were plated on the silicon nanowire substrates and cultured overnight.
The cells were removed by trypsin treatment and re-plated on a glass slide with lithographically defined strip lines for ESR measurements.
The samples were stained with calcein-AM and ethidium homodimer-1 for the live/dead assay.

%\end{methods}

\section*{Acknowledgements}
We thank R. Walsworth, V. Denic, C. Latta, L. Jiang, A. Gorshkov, P. Cappellaro, A. Sushkov and I. Lovchinsky for helpful discussions and experimental help. This work was supported by NSF, the Center for Ultracold Atoms, the Defense Advanced Research Projects Agency (QUASAR programs), Army Research Office (MURI program), NIH Pioneer Awards (5DP1OD003893-03) and NHGRI (1P50HG006193-01), the Swiss National Science Foundation (PBSKP2 \textunderscore 143918) (P.C.M.).
Correspondence and requests for materials should be addressed to HP (Hongkun\textunderscore Park@harvard.edu) and MDL (lukin@physics.harvard.edu)

\bibliography{biblatex-nature}

\newpage
\includegraphics[width=\textwidth]{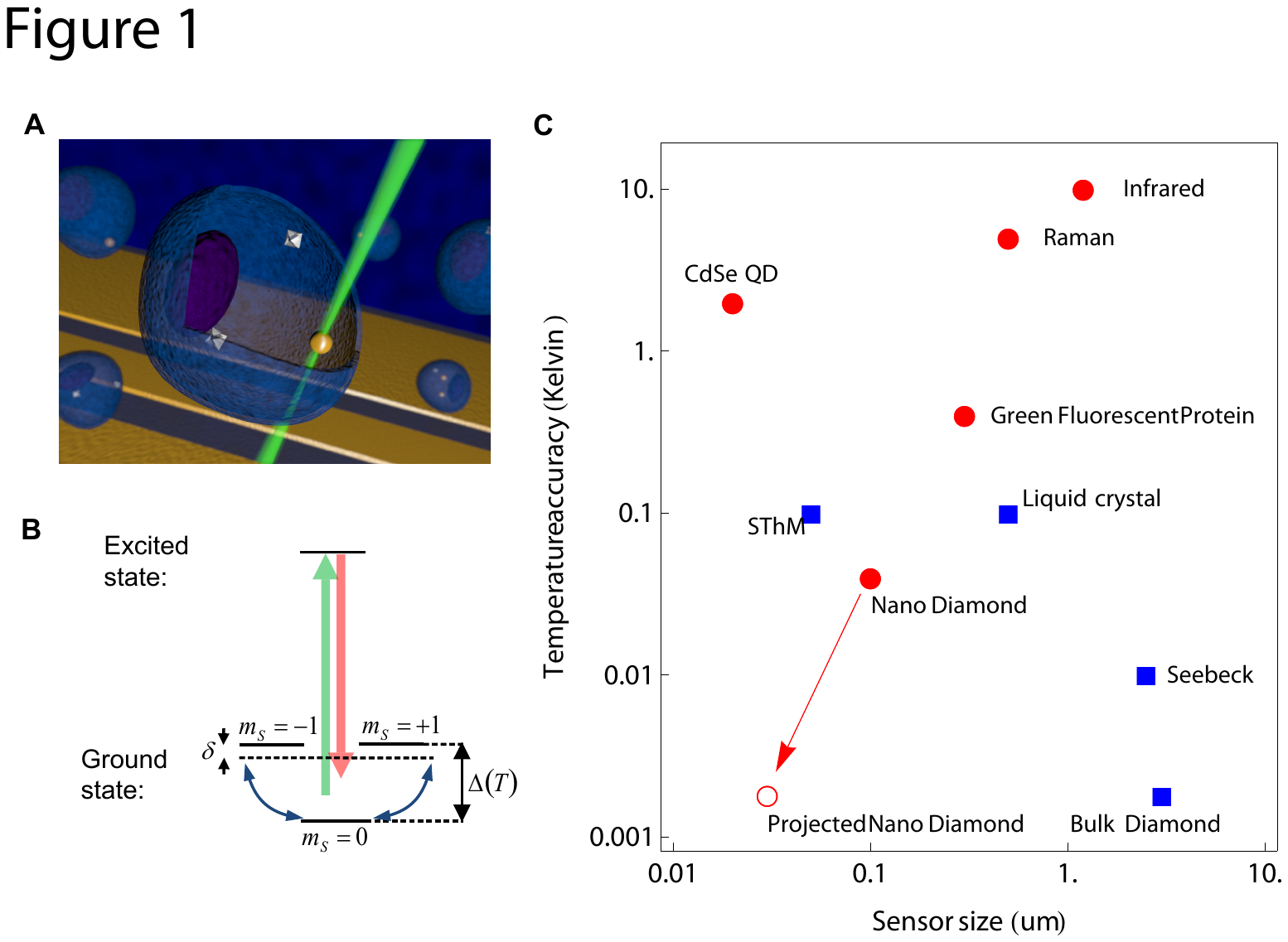}
\newpage
\includegraphics[width=\textwidth]{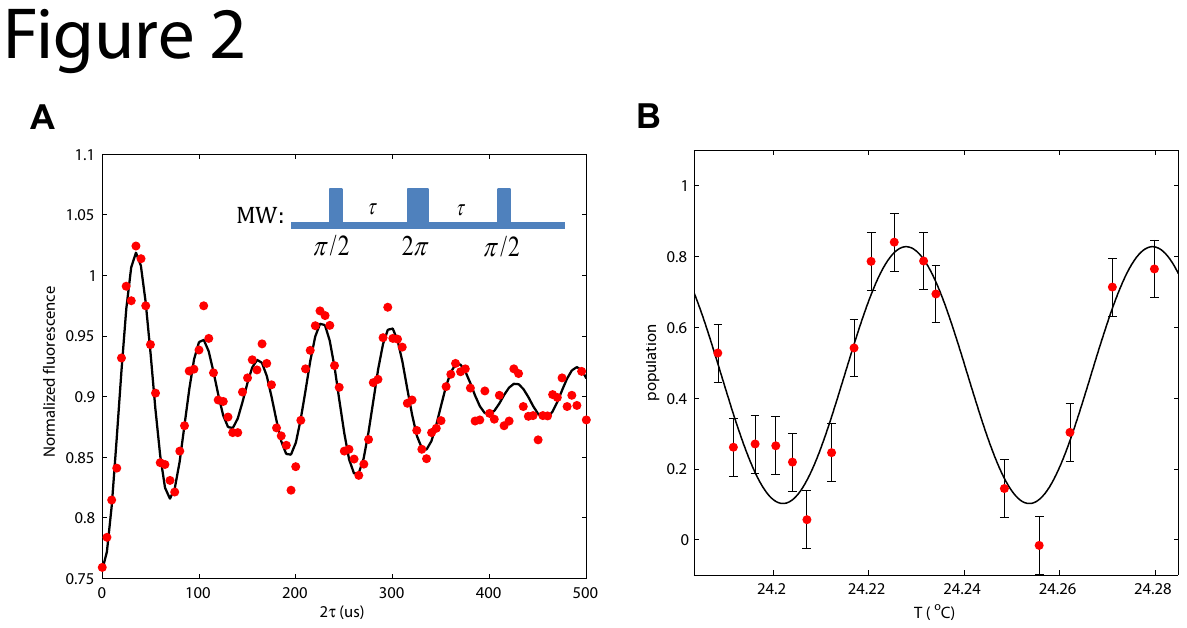}
\newpage
\includegraphics[width=\textwidth]{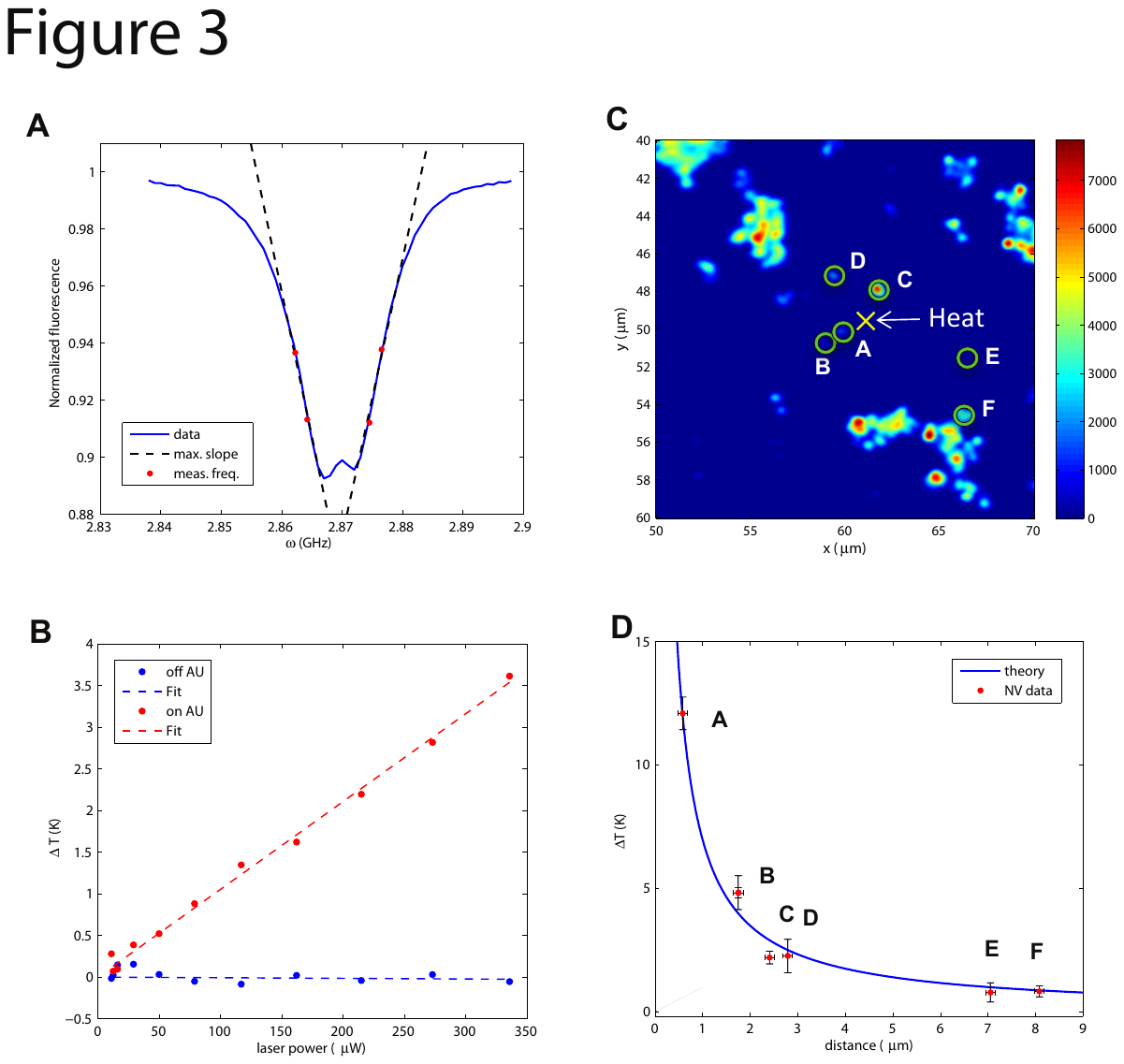}
\newpage
\includegraphics[width=\textwidth]{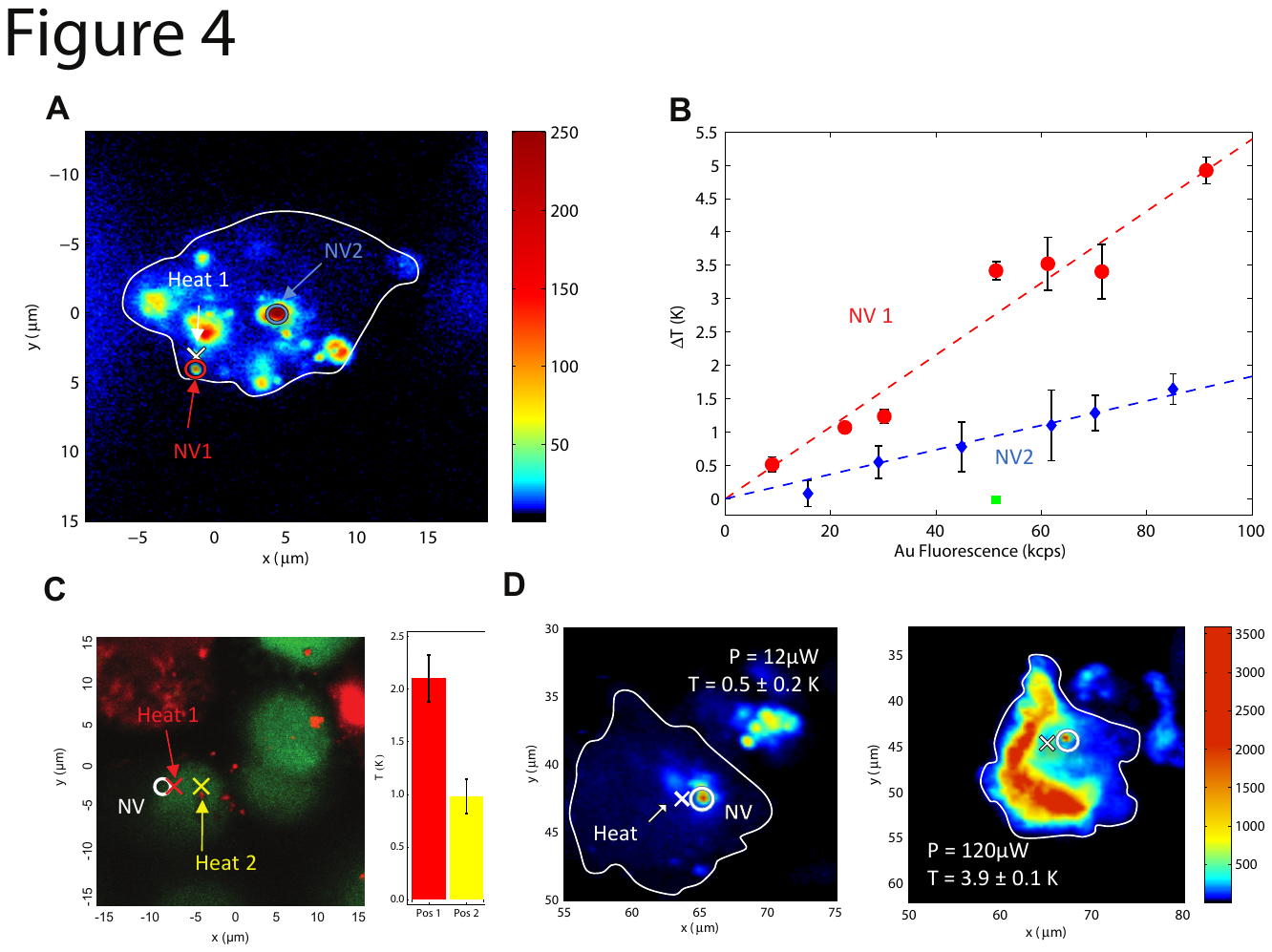}

\end{document}